\documentclass[%
reprint,
 amsmath,amssymb,
 aps,
floatfix,
]{revtex4-1}

\usepackage{graphicx}
\usepackage{dcolumn}
\usepackage{bm}

\usepackage{gensymb}
\usepackage{color}

\begin{document}

\preprint{APS/123-QED}

\title{Potential Energy Landscape of TIP4P/2005 water}

\author{Philip H. Handle}
\author{Francesco Sciortino}%
\affiliation{%
 Department of Physics\\
 Sapienza – University of Rome\\
 Piazzale Aldo Moro 5, I-00185 Roma, Italy
}%

\date{\today}

\begin{abstract}
We report an exhaustive numerical study of the statistical properties of the  potential energy landscape
of  TIP4P/2005, one of the most accurate rigid water models.
We show that, in the region where equilibrated configurations can be generated, 
  a  Gaussian landscape  description is able to properly describe the model properties.
We also find that  the volume dependence of the  landscape properties is consistent with the existence of a locus of density maxima in the phase diagram.  The  landscape-based  equation of state  accurately reproduces the  TIP4P/2005 pressure-vs-volume curves, providing a  sound  extrapolation of the free-energy 
at low $T$.   A positive-pressure liquid-liquid critical point is predicted by the resulting free-energy.
\end{abstract}

\pacs{Valid PACS appear here}
\maketitle


\section{Introduction}

The potential energy landscape (PEL) framework
 offers a  intuitive  description
of the physics of  low temperature ($T$) liquids.  It is based on the  idea that molecular motions at low $T$ can be split into   anharmonic vibrations around   potential energy local minima (the so-called inherent structures~\cite{stillinger1982hidden}, IS) and   infrequent visits of several different such minima~\cite{goldstein1969viscous}.  

Computer simulations have been crucial in supporting the PEL approach~\cite{stillinger1982hidden}. 
Efficient conjugate gradient minimization algorithms provide the possibility to
associate to each  equilibrium configuration its IS 
 and to study the connection between the system dynamics and
the  PEL~\cite{la2006relation,heuer2008exploring}.  The onset of a two step relaxation decay of correlation functions 
has been shown to coincide with the
onset of the PEL dominated region~\cite{sastry1998signatures}, i.e. 
with the temperature below which the  energy  $e_\text{IS}$  of the sampled IS  become $T$ dependent.  
As envisioned by Goldstein~\cite{goldstein1969viscous}, below this $T$  deeper and deeper PEL basins  
 are visited on supercooling.  

The power of the PEL approach is rooted not only in the possibility of closely
comparing the theoretical assumptions with numerical results, but more importantly in the
possibility of developing a  formal description of the thermodynamics of  supercooled liquids.
The number of  basins of depth $e_\text{IS}$  and  their shape  are key ingredients  to
express the liquid partition function (and hence the free energy) in terms of 
 statistical properties  of the 
 PEL~\cite{sciortino1999inherent,la2002potential,heuer2000density,debenedetti2001theory,debenedetti2003model,shell2004thermodynamics}.   Modelling of such quantities (supported by 
 a one-to-one comparison with numerical results) offers the possibility to predict, with 
 clear assumptions, the thermodynamics  of supercooled liquids~\cite{la2002potential} and in limited cases even in out-of-equilibrium conditions~\cite{sciortino2001extension}.
In addition, the analysis of the ISs provides insights in the glass phases of the material studied.

An interesting application of the PEL framework is offered by the study of water, a 
liquid which  continues to challenge contemporary science due to its complex behaviour~\cite{debenedetti03-jpcm,loerting11-pccp,gallo16-crev,handle2017supercooled}. A hallmark of this complexity is the well-known maximum in density at ambient pressure $(P)$ around 4~\degree C and the extrema
displayed by several thermodynamic response functions, as the isobaric heat capacity and the isothermal compressibility.  In addition, these response functions show a marked change in supercooled
states. For state of the art results see for example Refs.~\onlinecite{kim2017maxima,holten2017compressibility}.
Several recent reviews~\cite{loerting11-pccp,nilsson15-natcom,gallo16-crev,handle2017supercooled}  discuss in detail the  principal thermodynamic scenarios
compatible with the experimental observation: (i) the Speedy 
limit of stability scenario~\cite{speedy82-jpc}  (recently observed in numerical studies of colloidal model particles~\cite{rovigatti17-jcp,kalyuzhnyi2013two}) and (ii) the  LLCP scenario~\cite{poole92-nature}. 
This last scenario,
depending on the exact locus of the second critical point, changes into the singularity free scenario~\cite{sastry1996singularity} when the critical temperature approaches zero~\cite{stokely2010effect} or into the  critical point free scenario when the critical pressure approaches the spinodal pressure~\cite{anisimov2018thermodynamics}.

The statistical properties of the potential energy landscape (PEL) responsible for the density maxima and 
all other related~\cite{sastry1996singularity,rebelo1998singularity,poole2005density,holten2017compressibility} anomalies have been  previously discussed~\cite{sciortino03-prl}. As reviewed in Sec.\ref{sec:gpl}, within the
harmonic Gaussian PEL hypothesis, the volume dependence of just one of the  landscape 
parameters suffices to discriminate liquids with and without density anomalies.
An investigation of the PEL of the SPC/E model~\cite{spce} was shown to be consistent with 
theoretical predictions,   suggesting the presence of a low $T$ liquid-liquid critical point~\cite{sciortino03-prl}.  In more recent years, significantly improved classic rigid-water models have been proposed,
which are able to better reproduce water physical properties~\cite{vega11-pccp}.
Among this class of model potentials,  TIP4P/2005~\cite{abascal05-jcp} has emerged as the present-day optimal choice.  We present here the first detailed  potential energy landscape (PEL) investigation of the TIP4P/2005  model with the
aims of: (i) confirming the quality of the Gaussian PEL assumption in modelling the statistical
properties of the landscape; (ii) incorporating the anharmonic contribution to the basin shape which was
previously neglected; (iii) confirming the connection between statistical properties of the landscape and
density anomalies; (iv)  providing a sound PEL supported extrapolation of the model equation of state
(EOS) to investigate the possibility of a liquid-liquid critical point in  TIP4P/2005~\cite{abascal10-jcp,wikfeldt2011spatially,sumi13-rscadv,yagasaki14-pre, overduin13-jcp, overduin15-jcp,singh16-jcp,biddle17-jcp}.
In addition, the evaluation of the IS provides information on the structural properties of amorphous water~\cite{burton35-nature, bruggeller80-nature,mishima84-nature,loerting01-pccp,loerting11-pccp,handle15-pccp}.

\begin{figure*}[tb]
  \includegraphics[width=1\textwidth]{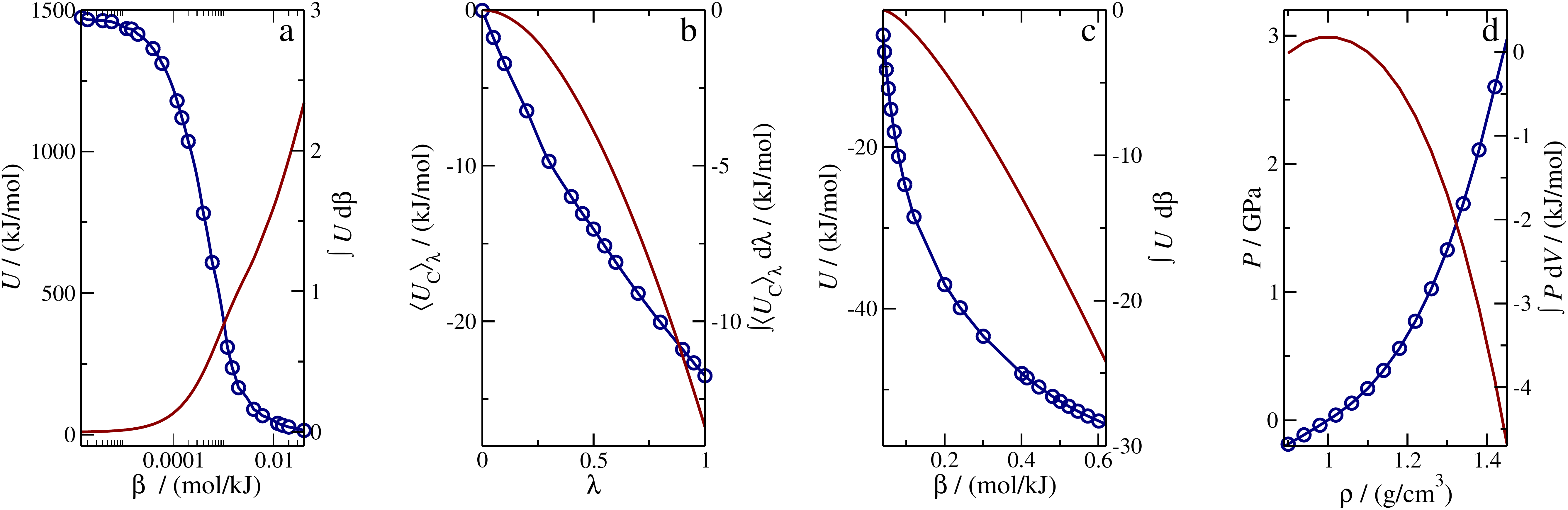}
\caption{Steps in the thermodynamic integration process  requested for the evaluation of the TIP4P/2005 free energy.
Panel  (a) shows the thermodynamic  integration from infinite $T$ (ideal gas) to $T_\text{ref}=3000$ K of  a system of particles  at $\rho_\text{ref}=1.1$ g/cm$^3$ interacting  according to the potential energy in Eq.~\ref{eq:ljm}.
Panel  (b) shows the Hamiltonian $\lambda$ integration from LJ to TIP4P/2005   (see Eq.~\ref{eq:lambda}).
Panel  (c) shows the thermodynamic  integration from $T_\text{ref}$ down to the studied $T$ at $\rho_\text{ref}$.   
Panel  (d) shows the thermodynamic  integration from $\rho_\text{ref}$  to the studied $\rho$  at $T=270$ K.   
In all panels, the blue line represents the integrand and 
the red line represents the running integral.
 }
\label{fig:ti}
\end{figure*} 

\section{The Gaussian PEL approach}
\label{sec:gpl}
We base our study on the potential energy landscape (PEL) framework, introduced by Stillinger and Weber~\cite{stillinger1982hidden,sciortino05-jsm}.
Within this framework, the multidimensional potential energy surface  $U(\vec{r}^{N},\phi^N,\theta^N,\psi^N)$, a function of the center of mass positions $\vec{r}$ and orientations (given by the Euler angles $\phi$, $\theta$, $\psi$) of all $N$ molecules, is split into \emph{basins}.  A basin is defined as the set of 
all configurations which under a steepest descent path end up in the same local potential energy minimum. Such minimum configuration  is named 
\emph{inherent structure} (IS) and its associated energy  $e_\text{IS}$. Thus,  $U(\vec{r}^{N},\phi^N,\theta^N,\psi^N)$ can be written as 
\begin{equation}
U(\vec{r}^{N},\phi^N,\theta^N,\psi^N) = e_\text{IS}+\Delta U(\vec{r}^{N},\phi^N,\theta^N,\psi^N),
\end{equation}
where $\Delta U(\vec{r}^{N},\phi^N,\theta^N,\psi^N)$ quantifies the energy associated to the thermal vibration  around the IS.
Grouping all basins with the same  $e_\text{IS}$,   the canonical partition function of the system can be written as~\cite{sciortino05-jsm,stillinger02-jcp}:
\begin{equation}
     Z(T,V)=\int_{e_\text{IS}}\Omega(e_\text{IS})\text{d}e_\text{IS}\ \text{e}^{- \beta F_\text{basin}(e_\text{IS},T,V)}
\end{equation}
where $\Omega(e_\text{IS})\text{d}e_\text{IS} $ is the number of basins with IS energy between  $e_\text{IS}$ and $e_\text{IS}+$d$e_\text{IS}$, $F_\text{basin}(e_\text{IS},T,V)$ is the average free energy of a basin of depth  $e_\text{IS}$, $\beta$ is $1/k_\text{B}T$ and  $k_\text{B}$  is  Boltzmann's constant.

A formal expression for the basin free energy can be written as 
\begin{align}
    F_\text{basin}(e_\text{IS},T,V)=&e_\text{IS}+F_\text{harm}(e_\text{IS},T,V)+
    \nonumber \\
     &+F_\text{anh}(e_\text{IS},T,V),
\end{align} 
\noindent
where the first term on the rhs  is the basin minimum energy, the second term accounts for the harmonic vibrations around the
minimum (and their $e_\text{IS}$ dependence) while the last term accounts for the remaining anharmonic contribution to the basin free energy.  
The harmonic free energy can be calculated as

\begin{equation}
\beta F_\text{harm}(e_\text{IS},T,V) \equiv  \left<\sum_{i=1}^{6N-3}\ln\left(\beta\hbar\omega_i(e_\text{IS})\right) \right>_{e_\text{IS}}
\label{eq:fah}
\end{equation}
\noindent
where $\omega_i(e_\text{IS})$ are the normal mode frequencies  and $\hbar$ is Planck's constant.
To separate the $T$ and the $e_\text{IS}$ dependence we write 
\begin{align}
\beta F_\text{harm}(e_\text{IS},T,V)  =&(6N-3)\ln\left(\beta A_0\right) +\nonumber \\  &+\mathcal S(e_\text{IS},V),
\end{align}
where 

\begin{equation}
  \mathcal S(e_\text{IS},V)=\left<\sum_{i=1}^{6N-3}\ln\left(\frac{\hbar\omega_i(e_\text{IS},V)}{A_0}\right) \right>_{e_\text{IS}}.
\label{eq:shape}  
\end{equation}
The latter is called the basin shape function and $A_0 \equiv 1$~kJ~mol$^{-1}$ ensures that the arguments of the logarithms bare no units.

The expressions derived so far are formally exact. To proceed further 
one needs to model the statistical properties of the landscape~\cite{heuer2000density,la2002potential,speedy2003energy,shell2004thermodynamics}
 as well as a description of the harmonic and anharmonic contributions. 
This is performed by comparing step by step the theoretical assumption with numerical results.

In the following we
review the equation of state (EOS) for a Gaussian landscape with minimal assumptions on the harmonic and anharmonic contributions and show
that the resulting EOS properly model the TIP4P/2005 pressure-volume relation. 
In a Gaussian landscape, for each $V$, three parameters describe  the PEL statistical properties: the total number of basins
 $\text{e}^{\alpha N}$ (where $N$ is the number of molecules), the most probable IS energy $E_0$ and the variance $\sigma^2$, resulting in

\begin{equation}
 \Omega(e_\text{IS})\text{d}e_\text{IS}=\frac{\text{e}^{{\alpha} N}}{\sqrt{2\pi{\sigma^2}}} \text{e}^{-\frac{\left(e_\text{IS}-{E_0}\right)^2}{2{\sigma^2}}}\text{d}e_\text{IS}.
\end{equation}

We further assume that (i)
 the shape function is linear with $e_\text{IS}$ (as previously found in several investigated models~\cite{giovambattista03-prl,sciortino03-prl,mossa2002dynamics}):
\begin{equation}
 \mathcal S(e_\text{IS},V) = a(V) + b(V) e_\text{IS},
 \label{eq:shape-fit}
\end{equation}
where $a$ and $b$ represent the ($V$-dependent) coefficients of the linear expansion;
(ii) that the anharmonic free energy is independent of $e_\text{IS}$.  Thus, we can write the anharmonic energy  as a polynomial in $T$ starting from a quadratic term
\begin{equation}
 E_\text{anh}(T,V)=\sum_{i=2}^{i_\text{max}}c_i(V) T^{i},
\label{eq:eanharm}
\end{equation}
where the $c_i$ represent the respective ($V$-dependent) coefficients. Solving $\text{d}S_\text{anh}/\text{d}E_\text{anh}=1/T$, 
the  anharmonic entropy is written as
\begin{equation}
 S_\text{anh}(T,V)=\sum_{i=2}^{i_\text{max}}\frac{i}{i-1}c_iT^{i-1}.
 \label{eq:sanh}
\end{equation}
Thus the anharmonic free energy is
\begin{align}
  F_\text{anh}(T,V)&=E_\text{anh}(T,V)-TS_\text{anh}(T,V)= \nonumber\\ 
&=\sum_{i=2}^{i_\text{max}}c_iT^{i}\left(1-\frac{i}{i-1}\right).\label{eq:fanh}
\end{align}
Other approximations,  which do not require the assumption of  $e_\text{IS}$ independence of the
anharmonic free energy
have been proposed in the past~\cite{lanave03-jpcm}, but
they do require a larger number of parameters.

Within the outlined approximation, the $T$ dependence  of the average IS energy $E_{\text{IS}}$ at a given $V$ can be formally written as

\begin{equation}
E_{\text{IS}}(T)=E_0-b\sigma^2-\frac{\sigma^2}{k_\text{B}T}, 
 \label{eq:eis-fit}
\end{equation}
where $E_0$, $\sigma^2$ and $b$ depend all on $V$.
Thus, in a Gaussian landscape $E_{\text{IS}}$ is linear in $1/T$, a prediction which can be tested numerically.

The configurational entropy can also be expressed in terms of $V$-dependent PEL quantities as
\begin{equation}
 \frac{S_\text{conf}}{k_\text{B}}\equiv  \ln \Omega(  E_{\text{IS}}(T))=
  \alpha N - \frac{\sigma^2(b+\beta)^2}{2}
 \label{eq:s-fit}
\end{equation}
which again provides a stringent numerical test of the $T$-dependence of $S_\text{conf}$.

Defining the  Kauzmann temperature  $T_\text{K}$ as the temperature at which 
 $S_\text{conf}=0$ one finds

\begin{equation}
k_BT_\text{K} = \left (\sqrt{ \frac{2\alpha N}{\sigma^2}} -b \right )^{-1}.
 \label{eq:tk}
\end{equation}
Below $T_K$ the system is  trapped in the basin of depth  
\begin{equation}
E_{\text{IS}}(T_\text{K})=E_0- \sqrt{2\alpha N} \sigma
\label{eq:ek}
\end{equation}



\begin{figure}[b]
  \includegraphics[width=0.5\textwidth]{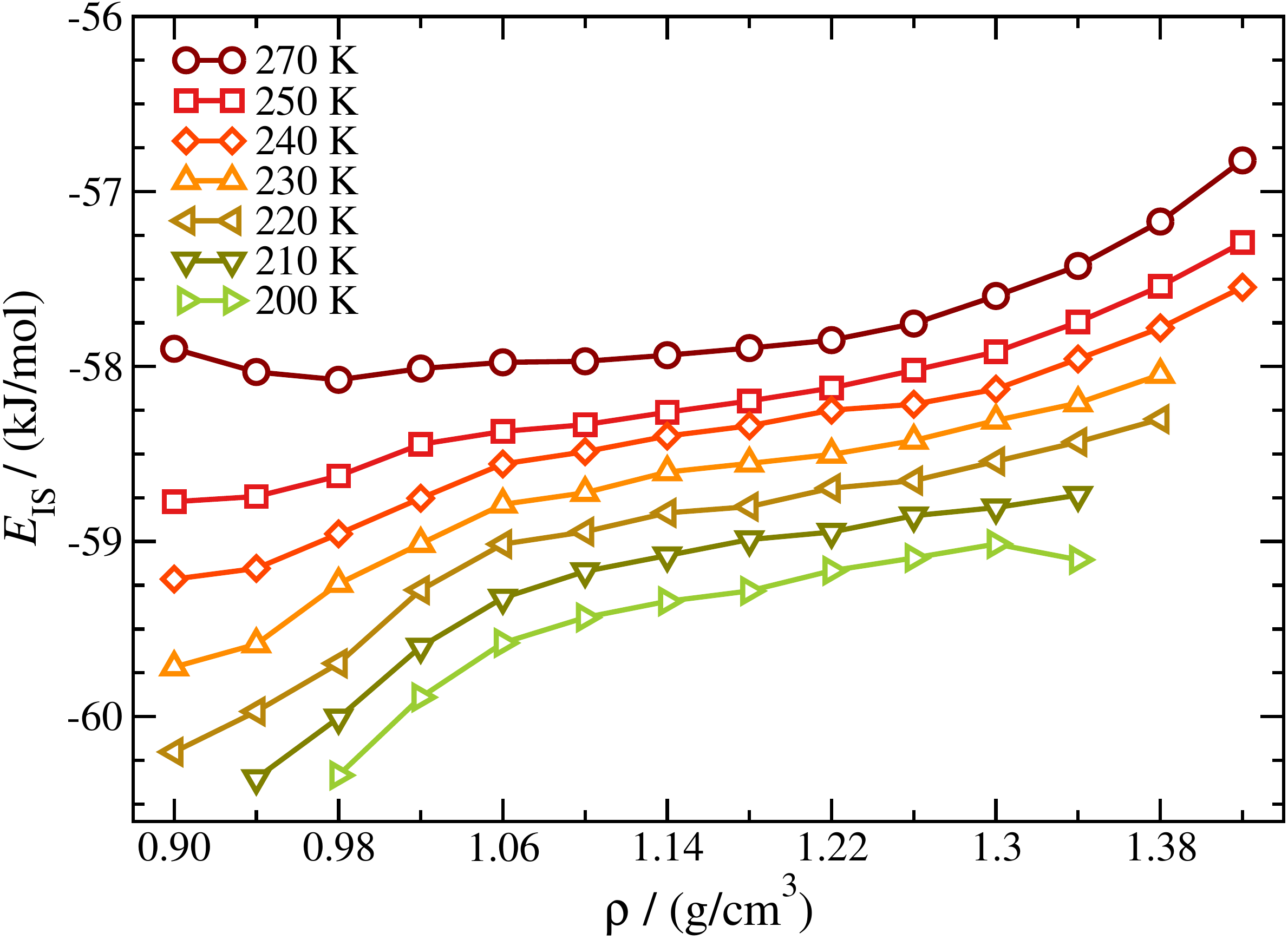}
\caption{Average inherent structure energy $E_\text{IS}$ for all studies state points.}
\label{fig:eis-vs-rho}
\end{figure} 

\begin{figure}[t]
  \includegraphics[width=0.5\textwidth]{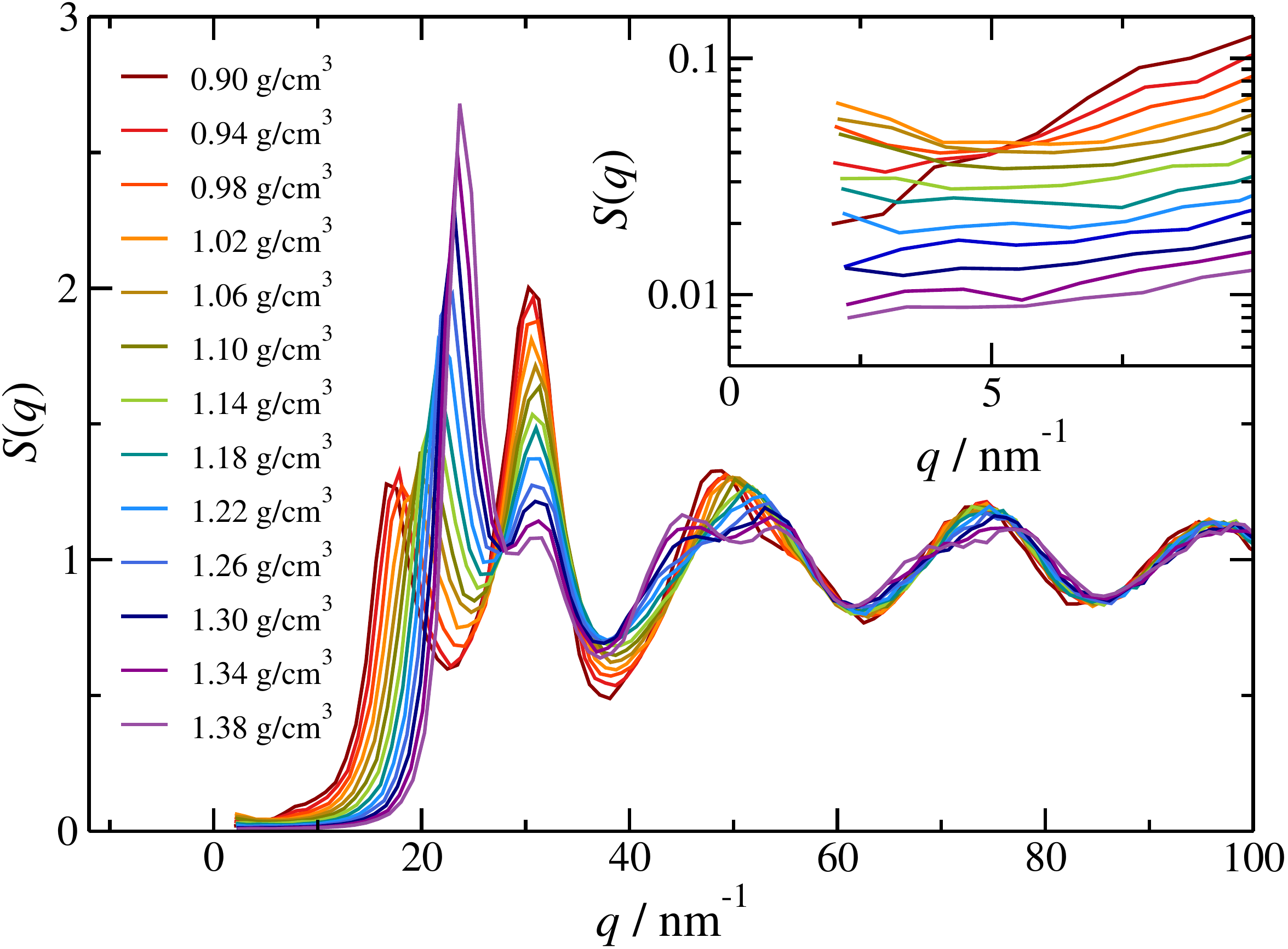}
\caption{Oxygen-oxygen structure factor evaluated in the IS at all studied densities at $T=220$ K. The inset enlarges the low $q$ limit.
 }
\label{fig:sq}
\end{figure} 

\section{The PEL Equation of State}

A benefit in using the Gaussian landscape approach lies in the possibility to analytically derive an equation of state (EOS). This EOS can be expressed in terms of volume derivatives of the PEL parameters~\cite{lanave02-prl,stillinger2002energy,shell2003energy}. Past studies however, used only the harmonic Gaussian parameters to formulate the EOS. We here derive an expression including anharmonic corrections. The PEL free energy
can be written as 
\begin{equation}
  F=  E_{\text{IS}}-TS_{\text{conf}}
  +F_{\text{harm}}+F_{\text{anh}}.
\end{equation}
The first two terms can be condensed in a free energy of the inherent structure $F_{\text{IS}}$.

Hence the pressure can be expressed as:
\begin{equation}
 P=-\frac{\partial F}{\partial V} =  -\frac{\partial F_{\text{IS}}}{\partial V}-\frac{\partial F_{\text{harm}}}{\partial V}-\frac{\partial F_{\text{anh}}}{\partial V}
\end{equation}
We now look at each term separately. The volume derivative of $F_{\text{IS}}$ is
\begin{align}
 -\frac{\partial F_{\text{IS}}}{\partial V} \nonumber
 &=-\frac{\partial }{\partial V}\left(E_\text{IS}-TS_\text{conf} \right)=\\
 &=-\frac{\partial (E_0-b\sigma^2)}{\partial V}+\frac{1}{k_\text{B}T}\frac{\partial \sigma^2}{\partial V}+T\frac{\partial S_\text{conf}}{\partial V}.\label{eq:dfisdv}
\end{align}
Using Equation~\ref{eq:s-fit} we further find:
\begin{align}
T\frac{\partial S_\text{conf}}{\partial V}
&=k_\text{B}T\frac{\partial}{\partial V}\left(\alpha N-\frac{\sigma^2\left(b+\beta\right)^2}{2}\right)=    \nonumber  \\
&=k_\text{B}T\frac{\partial}{\partial V}\left(\alpha N-\frac{b^2\sigma^2}{2}\right)-\frac{\partial(b\sigma^2)}{\partial V}-\frac{1}{2k_\text{B}T}\frac{\partial\sigma^2}{\partial V}\label{eq:dsdv}
\end{align}

Finally, bringing Eqn.~\ref{eq:dfisdv} and~\ref{eq:dsdv} together and grouping the terms  according to their respective temperature dependence
\begin{align}
-\frac{\partial F_{\text{IS}}}{\partial V} = &-\frac{\partial E_0}{\partial V}\nonumber \\
&+Tk_\text{B}\frac{\partial}{\partial V}\left(\alpha N-\frac{b^2\sigma^2}{2}\right)\nonumber \\
&+\frac{1}{T}\frac{1}{2k_\text{B}}\frac{\partial \sigma^2}{\partial V}
\end{align}

Next we perform the same procedure for $F_{\text{harm}}$ using Equations~\ref{eq:shape-fit} and \ref{eq:eis-fit} 

\begin{align}
-\frac{\partial F_{\text{harm}}}{\partial V} &=-k_\text{B}T\frac{\partial\mathcal S}{\partial V}=  \nonumber \\
&=-k_\text{B}T\frac{\partial}{\partial V}\left(a+bE_0-b^2\sigma^2-\frac{b\sigma^2}{k_\text{B}T}\right)
\end{align}

Sorting the terms according to their respective temperature dependence we find:

\begin{align}
-\frac{\partial F_{\text{harm}}}{\partial V} =& \frac{\partial}{\partial V}b\sigma^2 \nonumber \\
&-Tk_\text{B}\frac{\partial}{\partial V}(a+bE_0-b^2\sigma^2)
\end{align}

The anharmonic contribution to the pressure arises from the volume derivative of the $c_i$ coefficient in Eq.~\ref{eq:fanh}
\begin{equation}
-\frac{\partial F_\text{anh}}{\partial V} =
\sum_{i=2}^{i_\text{max}}T^i \frac{\partial c_i(V)}{\partial V}\left(\frac{i}{i-1}-1\right)
\end{equation}

Since all contributions could be separated into terms of different temperature dependence the equation of state including anharmonic corrections reads:
\begin{equation}
 P(T,V)=\sum_{i=-1}^{i_\text{max}}\mathcal P_{T^i}(V)T^i,
 \label{eq:eos}
 \end{equation}
 where we have defined 
 \begin{align}
 \mathcal P_{T^{-1}}(V)&= \frac{1}{2k_B}\frac{\text{d}}{\text{d} V } \sigma^2
 \label{eq:ptm1}\\
  \mathcal P_{T^0}(V)&=-\frac{\text{d}}{ \text{d} V } \left(E_0-b\sigma^2\right)
 \label{eq:pt0}\\
    \mathcal P_{T^{1}}(V)&= k_B\frac{\text{d}}{\text{d} V }\left(\alpha N -a -bE_0+\frac{b^2\sigma^2}{2}\right)
 \label{eq:pt1}\\
 \mathcal P_{T^{i\geq2}}(V)&=\left(\frac{i}{i-1}-1\right)\frac{\text{d} }{\text{d} V}c_i(V).
 \label{eq:pti}
 \end{align}
For reasons that will  be clear next, we notice that $\mathcal P_{T^{-1}}(V)$ involves only the $V$ derivative of $\sigma^2$.

From the mathematical structure of Eqn.~\ref{eq:eos} it is clear that, if the system moves along an isochore, the high $T$ behaviour is controlled by 
the $i_\text{max}$ order term in T. The behaviour at low $T$ however is controlled by the $\mathcal P_{T^{-1}} T^{-1}$ term.
One can also see that the pressure along an isochore must display a minimum  if $\mathcal P_{T^{-1}}> 0$. In this  case,
a density maximum  exists.  Indeed, according to a Maxwell relation,   $\partial P/\partial T |_V = 0$ corresponds to  $\partial V/\partial T |_P = 0$, i.e. to a density extremum.
Hence, the condition for the existence of density maxima ($P_{T^{-1}}> 0$)  in the PEL formalism, corresponds  to $\text{d}\sigma^2/\text{d}V > 0$ (see Eq.~\ref{eq:ptm1}).
Thus, in the Gaussian landscape,  liquids with density anomalies 
must be characterized by a $V$-range where $\sigma^2$ increases with $V$.

\begin{figure*}
  \includegraphics[width=1\textwidth]{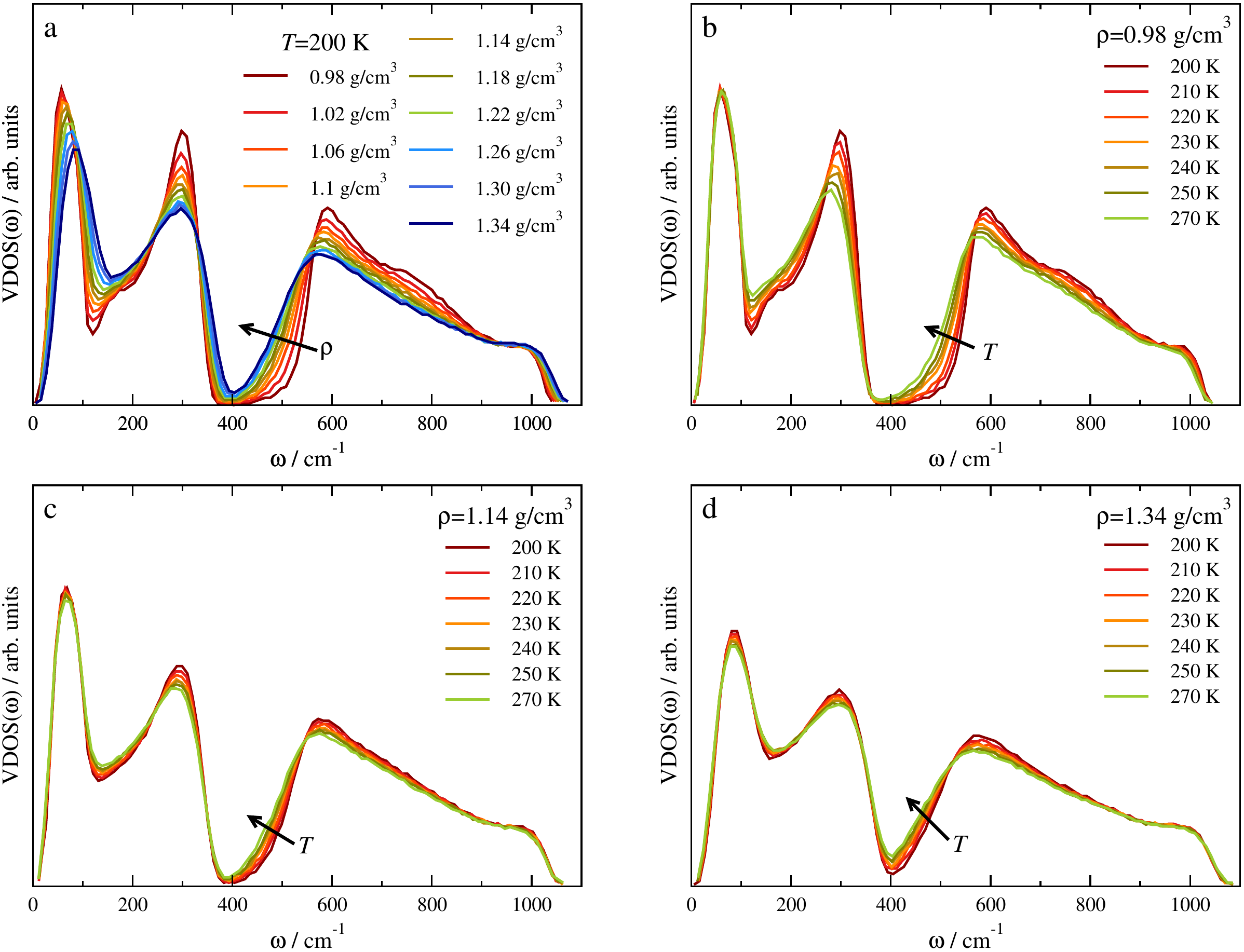}
\caption{  Vibrational density of states resulting from the diagonalization of the Hessian matrix. Panel (a) shows the density dependence at $T=200$ K. Panel (c), (d) and (e)
report the $T$ dependence at $\rho=0.98$, $\rho=1.14$ and $\rho=1.34$ g/cm$^3$ respectively. 
 }
\label{fig:vdos}
\end{figure*} 

\section{Simulation Details}

\subsection{NVT simulations}
We perform NVT simulations of 1000 TIP4P/2005 molecules in a cubic box utilising GROMACS 5.1.2~\cite{vanderspoel05-jcc} with a leap-frog
  integrator using a timestep of 1~fs. The temperature is controlled using a  Nos\'{e}-Hoover thermostat~\cite{nose84-mp,hoover85-pra} with a time constant of 0.2~ps. For the coulombic interactions we use a particle mesh Ewald treatment~\cite{essmann95-jcp} with a Fourier spacing of 0.1~nm. For both the Lennard-Jones and the real space Coulomb interactions, a cut-off of 0.85~nm is used.   Lennard-Jones interactions beyond 0.85~nm have been included assuming a uniform fluid density.
  Finally, we maintain the bond constraints using the LINCS (Linear Constraint Solver) algorithm~\cite{hess08-jctp} of 6$\text{th}$ order with one iteration to correct for rotational lengthening. We investigate 14 different densities from 0.9 to 1.42~g/cm$^{3}$ and seven different $T$s between 200 and 270~K. Very long equilibration runs (up to  100 ns)  followed by equally long production runs have  been performed. Equally spaced configurations from the production runs have been used 
  in the following analysis.
  
\subsection{Inherent Structures}

To generate the IS configuration we minimise the potential energy of the system  with a conjugate gradient method (evolving the centre of mass and the
orientation of the particles around their principal axes). 
At least 30 configurations  extracted from each trajectory were minimised. 
We also evaluate  the normal modes of all the found inherent structures \emph{via} the numerical determination of the Hessian, the $6N \times 6N$ matrix of the
second derivatives of the potential energy as a function  of the molecule centre of mass and  principal axes.  Both conjugate gradient minimization and
Hessian evaluation has been performed with our own code. 

 \begin{figure}
  \includegraphics[width=0.5\textwidth]{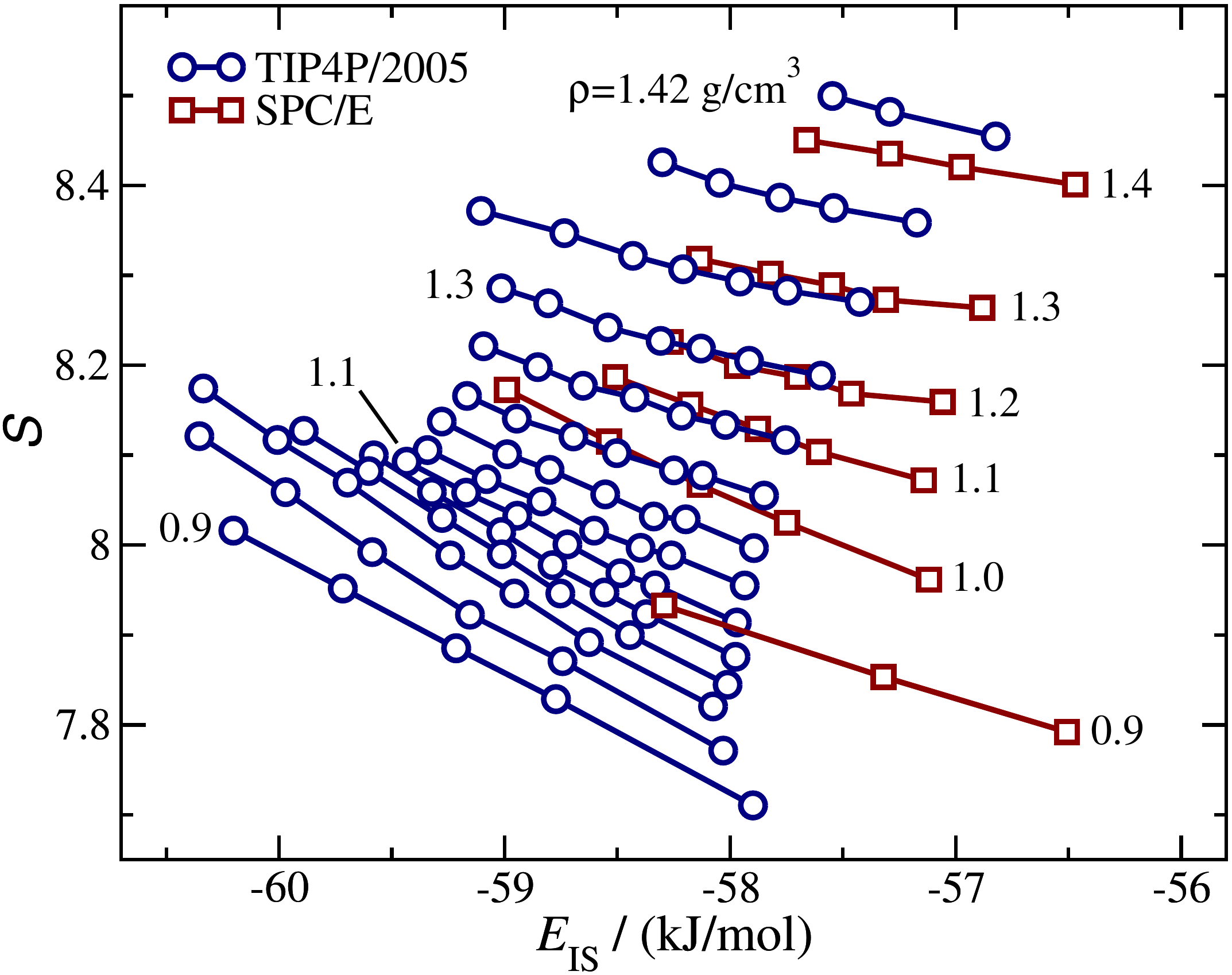}
\caption{$E_\text{IS}$ dependence of the shape function $\mathcal{S}$  (per particle) for all studied densities (blue circles) in comparison with results for SPC/E from Ref.~\cite{giovambattista03-prl} (red squares).}
\label{fig:shape}
\end{figure} 

\begin{figure}
  \includegraphics[width=0.5\textwidth]{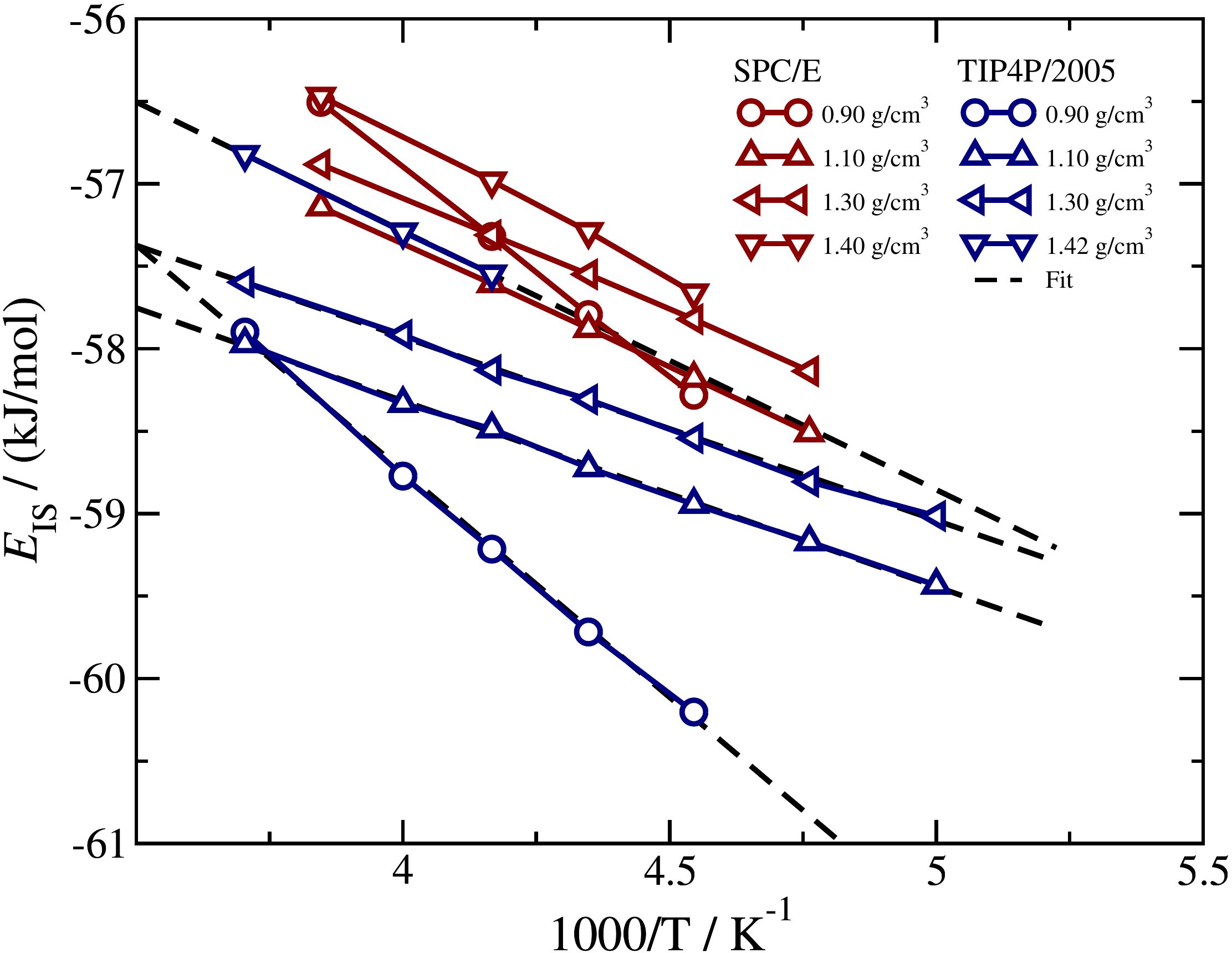}
\caption{Inverse temperature dependence  of   $E_\text{IS}$ for selected densities   (blue data, black dashed fits) in comparison with results for SPC/E from Ref.~\cite{lanave04-jpcb} at similar densities (red).}
\label{fig:eis-t}
\end{figure} 

\subsection{Free energy}
\label{sec:free}

We evaluate the harmonic free energy ($F_\text{harm}$) of all different IS according to Eq.~\ref{eq:fah}, starting form the  density of states  calculated by diagonalising the Hessian matrix.
Instead, 
we evaluate the  free energy of  TIP4P/2005 in the liquid state $(F_{\text{liq}})$
 performing  thermodynamic and Hamiltonian integration~\cite{vega08-jpcm} starting from the known reference free energy
of an ideal gas of water-shaped molecules $F_\text{id}(T,V,N)$. 
Specifically,
\begin{equation}
\beta F_\text{id}(T,V,N)=-\ln Z_\text{id}(T,V,N)
\end{equation}
where the partition function
\begin{equation}
Z_\text{id}(T,V,N)=\frac{Z_\text{T} Z_\text{R}}{N!},
\end{equation}
can be split in a translational part
\begin{equation}
 Z_\text{T}=\left(V\left(\frac{2\pi mk_\text{B}T}{h^2}\right)^\frac{3}{2}\right)^N
\end{equation}
and a rotational part
\begin{equation}
 Z_\text{R}=\left(\frac{1}{2}\left(\frac{8\pi^2k_\text{B}T}{h^2}\right)^\frac{3}{2}\left(\pi I_xI_yI_z\right)^\frac{1}{2}\right)^N,
\end{equation}
where $m$ is the mass of the water molecule and $I_x$, $I_y$ and $I_z$ are the moments of inertia along the three principal axes. 
The factor $\frac{1}{2}$ in front of $Z_\text{R}$ accounts for the water molecule's $C_{2v}$ symmetry~\cite{mayer1963}.
For future reference, we notice that the molecular ideal gas (non-interacting) free energy at $T_\text{ref}=3000$~K
and   $\rho_\text{ref}=1.1 $ g/cm$^3$  is $-481.15$~kJ/mol.

To evaluate the free energy of a system of water molecules (e.g. with centre of mass and orientational degrees of freedom) but interacting only via a 
Lennard-Jones interaction $U_\text{LJ}(r)$ between the oxygen sites (we select the same $\sigma=0.31589$ nm and the same $\epsilon=774.9$~J~mol$^{-1}$
of the TIP4P/2005 model~\cite{abascal05-jcp}) 
we perform a thermodynamic
integration along a path of constant  density $\rho_\text{ref}$ from infinite $T$ down to 
$T_\text{ref}$ of the isotropic pair potential $U(r)$ defined as~\cite{smallenburg2015phase} 

\begin{equation}
U(r) = \text{min}(U_\text{LJ}(r), U_\text{cutoff})
\label{eq:ljm}
\end{equation}

This potential coincides with the $U_\text{LJ}(r)$  potential for all
intermolecular distances for which  $U_\text{LJ}(r)<U_\text{cutoff}$ and   
it is constant and equal to $U_\text{cutoff}$  otherwise. 
With this choice, the divergence
of the potential energy for configurations in which some intermolecular separations vanish (which would otherwise
be probed at very high $T$) is eliminated and the infinite $T$ limit is properly approximated
by an ideal gas of molecules at the same density.  Specifically, we choose $U_\text{cutoff}=100 RT_\text{ref}$ J/mol ($R$ being the ideal gas constant), corresponding to a interparticle distance $r_\text{cutoff}= 0.18033$ nm.

The fluid free energy (per particle) is calculated as
\begin{equation}
F_\text{LJ}(T,V,N)= F_\text{id}(T,V,N) + k_\text{B} T_\text{ref} \int_0^{\beta_\text{ref}}  \left< U \right>_\beta \text{d}\beta
\end{equation}
where the integration goes from infinite $T$ to $\beta_\text{ref}=1/k_\text{B} T_\text{ref}$. Fig.~\ref{fig:ti}(a) shows the resulting $\beta$ dependence of  $\left< U \right>_\beta$
evaluated on a mesh of 23 points  and the corresponding running integral  $\int_0^\beta  \left< U \right>_\beta \text{d}\beta$. The free energy of water-shaped molecules intarcting only \emph{via} a Lennard-Jones interaction at $T_\text{ref}$ and  $\rho_\text{ref}$  is $-422.84$ kJ/mol.

Next we evaluated the free energy change from the LJ to the TIP4P/2005 model at $\rho_\text{ref}$ and $T_\text{ref}$ via Hamiltonian integration 
interpolating from  LJ  to  TIP4P/2005. Hence we perform simulations  based on the 
potential  energy $U_\text{LJ} + \lambda (U_\text{TIP4P/2005}-U_\text{LJ})$ for 15 different $\lambda$  values. In this way the
electrostatic interactions are progressively turned on.  
The resulting  TIP4P/2005 free energy ($ F_\text{liq}$) can be calculated as 

\begin{equation}
 F_{\text{liq}}(T,V,N)= F_\text{LJ}(T,V,N) + \int_0^1  \left< U_\text{C}\right>_\lambda \text{d}\lambda,
\label{eq:lambda}
\end{equation}
\noindent
where $\left< U_\text{C}\right>_\lambda$ is the canonical average of the potential energy difference $U_\text{TIP4P/2005}-U_\text{LJ}$ evaluated in a simulation with 
potential energy $U_\text{LJ} + \lambda (U_\text{TIP4P/2005}-U_\text{LJ})$.   
Fig.~\ref{fig:ti}(b)  shows $\left< U_\text{C}\right>_\lambda$ and the corresponding integral. As a result of the integration  we estimate  
$F_\text{liq}(\rho_\text{ref},T_\text{ref})=-436.23$~kJ~mol$^{-1}$.
From this reference point, we then calculate 
 via standard thermodynamic integration along isochores (see Fig.~\ref{fig:ti}(c) for $\rho_\text{ref}$)
   and/or along isotherms (see Fig.~\ref{fig:ti}(d) for $T=270$ K)  the TIP4P/2005 free energy at any $T$ and $\rho$.  As a reference for future studies we report the resulting
     free-energy in the range $270-200$ K for all investigated densities in  Appendix A.

From the free energy $F_\text{liq}$ and the total energy of the liquid $E_\text{liq}$ the entropy can be calculated from
\begin{equation}
 S_\text{liq}=\frac{E_\text{liq}-F_\text{liq}}{T}
\label{eq:sliq} 
\end{equation}

\begin{figure*}
  \includegraphics[width=1\textwidth]{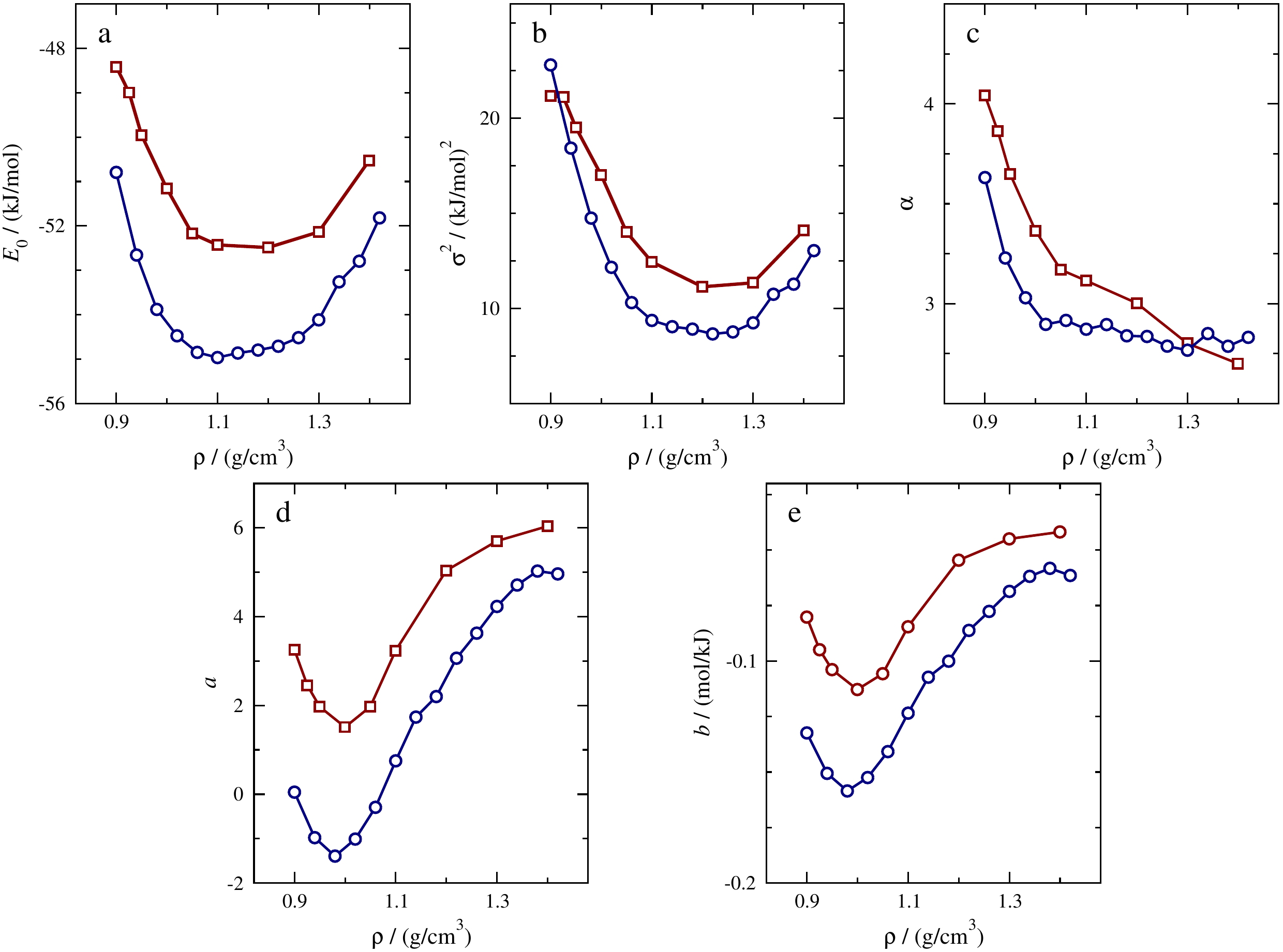}
\caption{Density dependence of the landscape parameters $E_0$, $\sigma^2$, $\alpha$, $a$ and $b$ in comparison with corresponding data for SPC/E from Ref.~\cite{sciortino03-prl}.}
\label{fig:params}
\end{figure*}

\section{Results}

\subsection{Minima of the TIP4P/2005}

In this section we provide information on the structure of the IS, describing  the IS energies, the 
static structure factor between the oxygen centres $S(q)$ and  the local curvature of the potential energy around the
IS (i.e. the vibrational density of states).   Since the IS are essentially realisations of the possible glasses of a material,
the following structural data provide information on the possible structure and vibrational dynamics of water glasses. 

Fig.~\ref{fig:eis-vs-rho} shows the calculated IS energies for all studied state points.  As expected, deeper and deeper basins are
explored on cooling.  The deepest IS are sampled at the lowest explored density $(\rho \approx 0.9$ g/cm$^3$),  revealing the
optimal density for the build up of the hydrogen-bonded network.  Interestingly, a region of negative curvature
of $E_\text{IS}$ vs $\rho$ is present at low $T$, a first indication of an energetic destabilisation of the liquid~\cite{sciortino1997line,giovambattista2017influence}.

Fig.~\ref{fig:sq} shows  $S(q)$  for different densities at $T=220$ K.   Being evaluated in the local minimum  the IS $S(q)$ 
reflects the static correlation present in the IS, in the absence of any thermal broadening. A clear progression of the
structure on increasing density is observed.  The signature of the tetrahedral ordering, which show in the 
$S(q)$ as a pre-peak around 17 nm$^{-1}$ and a main peak around 31 nm$^{-1}$, progressively disappear in favour of a
main peak around 25 nm$^{-1}$ at densities so high that the hydrogen bond network is strongly perturbed. 
The inset shows the low $q$ region, to provide a quantification of the system compressibility (related to $S(0)$).
A clear maximum in the density dependence of $S(q)$ at the smallest accessible $q$ value appears, signalling the
presence of an extremum in the structural component  of the thermal  compressibility.   The density fluctuations 
implicit in the structure achieve their maximum value when $\rho\approx 1.02$ g/cm$^3$.
We also note  (see inset of Fig.~\ref{fig:sq}) that  the extrapolation of $S(q)$ at vanishing wavevectors reaches values  of the order of $10^{-2}$ for both the network density ($\rho=0.9$)  and for the highest studied density ($\rho=1.42$ g/cm$^3$).  The vanishing of $S(0)$ in disordered systems has been interpreted as evidence
of hyper-uniformity~\cite{torquato2003local}. Recently, it has also been suggested that the structure factor of amorphous ices  
shows strong signatures of hyper-uniformity~\cite{PhysRevLett.119.136002}.

Fig.~\ref{fig:vdos}  shows the vibrational density of states (VDOS), in harmonic approximation, for several $\rho$ and $T$,  resulting from diagonalising the Hessian matrix in the IS. 
Previous evaluation of the TIP4P-2005 VDOS in a limited frequency range had been based on Fourier transform of the oxygen velocity autocorrelation functions~\cite{kumar2013boson}. The VDOS enters in the evaluation of the basin harmonic vibrational entropy. 
In all cases, a clear separation between
the low-frequency translational bonds  ($\omega < 400 $ cm$^{-1}$) and the higher frequency librational bands  ($\omega > 400 $ cm$^{-1}$)  is observed. 
  Fig.~\ref{fig:vdos}  shows that the $T$ dependence of the VDOS 
is more significant at $\rho=0.98$ g/cm$^3$, where the development of the tetrahedral networks takes place on cooling and less significant
at $\rho=1.34$ g/cm$^3$, where the structure of the system is less dependent on the hydrogen bond formation, as clearly indicated by $S(q)$. 
The data in Fig.~\ref{fig:vdos}(a) show that instead a strong density dependence is observed at low $T$.   The low frequency part of the VDOS
behaves as $\omega^2$ as expected in the Debye limit.

To evaluate the $e_\text{IS}$ dependence of the basin shape in harmonic approximation we evaluate the 
 function $\mathcal{S}$ (cf. Equation~\ref{eq:shape}).  
  For all densities,   $\mathcal{S}$  is    linear 
 in $e_\text{IS}$ (Fig.~\ref{fig:shape}), providing a simple quantification of the basin dependence of the free energy  (Eqn.~\ref{eq:shape-fit}) 
 via the intercept $a$ and slope $b$.  
 The basin shape  for TIP4P/2005 is  similar to the results for SPC/E~\cite{giovambattista03-prl} reproduced also in Figure~\ref{fig:shape}. The TIP4P/2005 shape is only shifted down in $\mathcal{S}$ and $e_\text{IS}$. This similarity is also reflected in the fitting parameters $a$ and $b$ for both models 
 as will be shown below.

\subsection{Evaluating the PEL parameters}

Figure~\ref{fig:eis-t} shows $E_\text{IS}$ as a function of $1/T$ for TIP4P/2005. This figure, and the following,  also show 
corresponding results for the SPC/E model of water, the only other water model for which a detailed PEL study has been previously performed~\cite{lanave04-jpcb}.
At all densities, $E_\text{IS}$ is well described by a linear  $1/T$  dependence, consistent with the predictions of 
 Eqn.~\ref{eq:eis-fit}. This supports the assumption of a Gaussian landscape description of the statistical properties of the PEL and provides a
straightforward measure of   the important landscape parameter $\sigma^2$. 
The TIP4P/2005 data  are consistently  smaller than the SPC/E data, but the overall trend of the two models is very similar (e.g., the curve at $\rho=0.9$~g/cm$^{3}$ is the steepest in both cases).

From the value of the parameters obtained performing linear representation of the data 
 reported in Figs.~ \ref{fig:shape} and~\ref{fig:eis-t} the density (or volume) dependence of the Gaussian PEL parameters $\sigma^2$ (Eq.~\ref{eq:eis-fit}) 
  and $E_0$ (Eq.~\ref{eq:eis-fit}) can be evaluated. The results are shown   in Fig.~\ref{fig:params}.  $E_0$ shows the expected minimum,
  indicating an optimal density ($\rho\approx 1.1$ g/cm$^3$) for energetic stabilisation, resulting from the compensation between the repulsive contributions (relevant at high densities)  and weakening of the attraction on stretching for low densities.  More interesting is the minimum observed in the $\rho$ (or $V$) dependence of 
  $\sigma^2$ which, as previously discussed, provides the landscape signature of anomalous behaviour.   The data shown in Fig.~\ref{fig:params}(b)
  shows that density anomalies are expected in the range $0.9<\rho<1.2$ g/cm$^3$, where $\text{d}\sigma^2/\text{d}V$ is  positive.
Fig.~\ref{fig:params}(d-e) show respectively the $\rho$ dependence of the linear fit of the shape ${\cal S}$ function.  
  
 To calculate  $\alpha$ we exploit Eq.~\ref{eq:s-fit}, and evaluate $\alpha$ as difference between
  $S_\text{conf}$ and a quantity dependent only on the previously calculated $b$ and $\sigma^2$.   $S_\text{conf}$ is the difference between
  the  entropy of the liquid $S_\text{liq}$ and the vibrational entropy of the explored basins $S_\text{harm}+S_\text{anh}$
   (harmonic and anharmonic vibrations around the inherent structures)
  and can be thus written as
\begin{equation}
 S_\text{conf}=S_\text{liq}-S_\text{harm}-S_\text{anh}.
 \label{eq:sconf}
\end{equation}

\begin{figure}[tb]
  \includegraphics[width=0.5\textwidth]{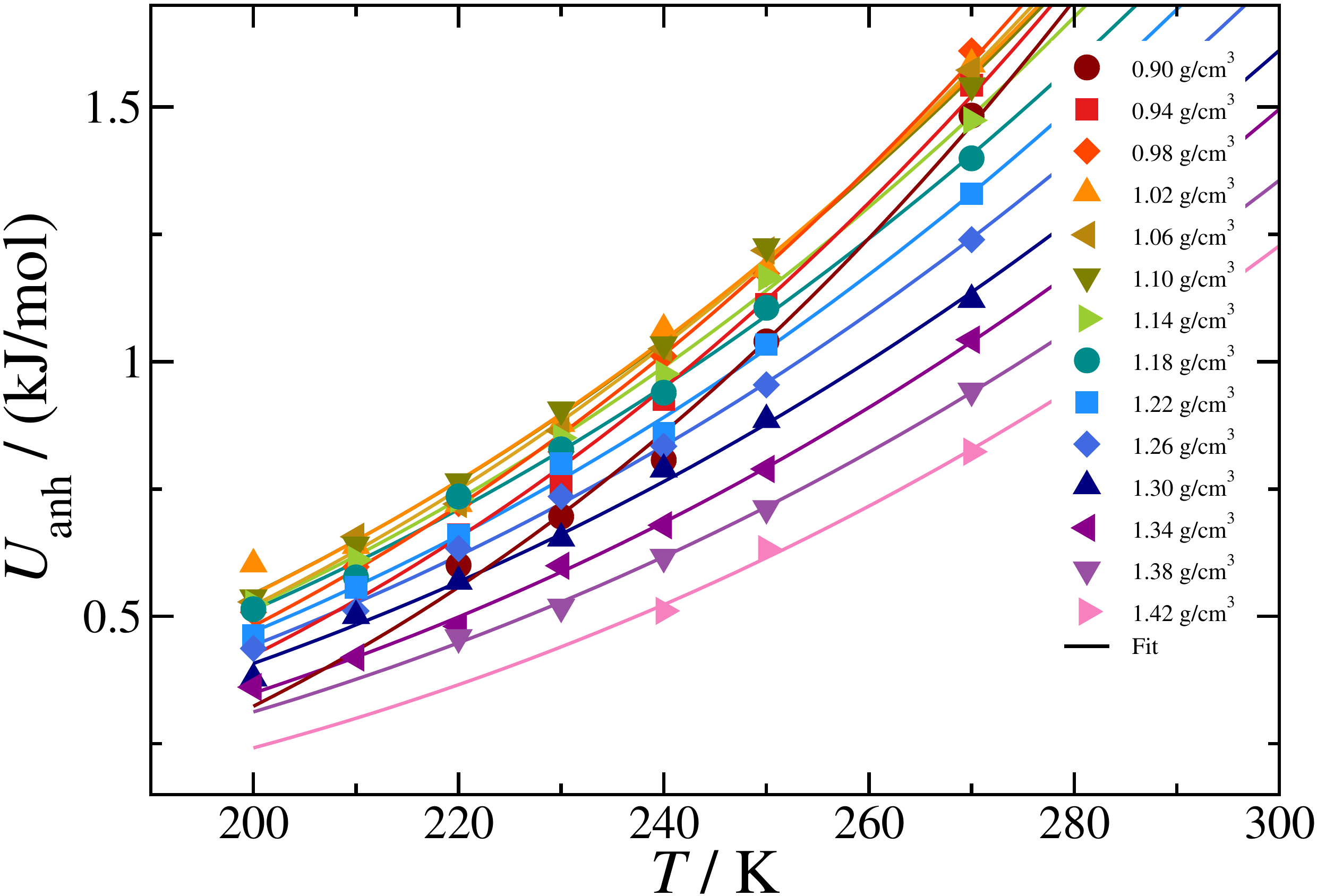}
\caption{Temperature dependence of the anharmonic potential energy for several different densities and the
associated polynomial fit $c_2(V) T^2 + c_3(V) T^3$, see Eq.~\ref{eq:eanharm}. }
\label{fig:uanh}
\end{figure}

All three contributions on the rhs are available: $S_\text{liq}$ from Eq.~\ref{eq:sliq}. 
$S_\text{harm}$ can be evaluated subtracting from $F_\text{harm}$  (Eq.~\ref{eq:fah}) the harmonic potential energy $(6N-3)k_\text{B}T/2$, and
  $S_\text{anh}$ according to Eqn.~\ref{eq:sanh}.  The evaluation of  $S_\text{anh}$  requires the preliminar modelling of
   the $T$ dependence of the anharmonic energy $E_\text{anh} \equiv U_\text{TIP4P/2005}-E_{\text{IS}}-(6N-3)k_\text{B}T/2$.    Fig.~\ref{fig:uanh}  shows 
  that   $E_{\text{anh}}$ vs. T is well represented by  Eq.~\ref{eq:eanharm}  with $i_\text{max}=3$ (e.g. $E_\text{anh}=c_2(V) T^2 + c_3(V) T^3$). 
 
\begin{figure}
  \includegraphics[width=0.5\textwidth]{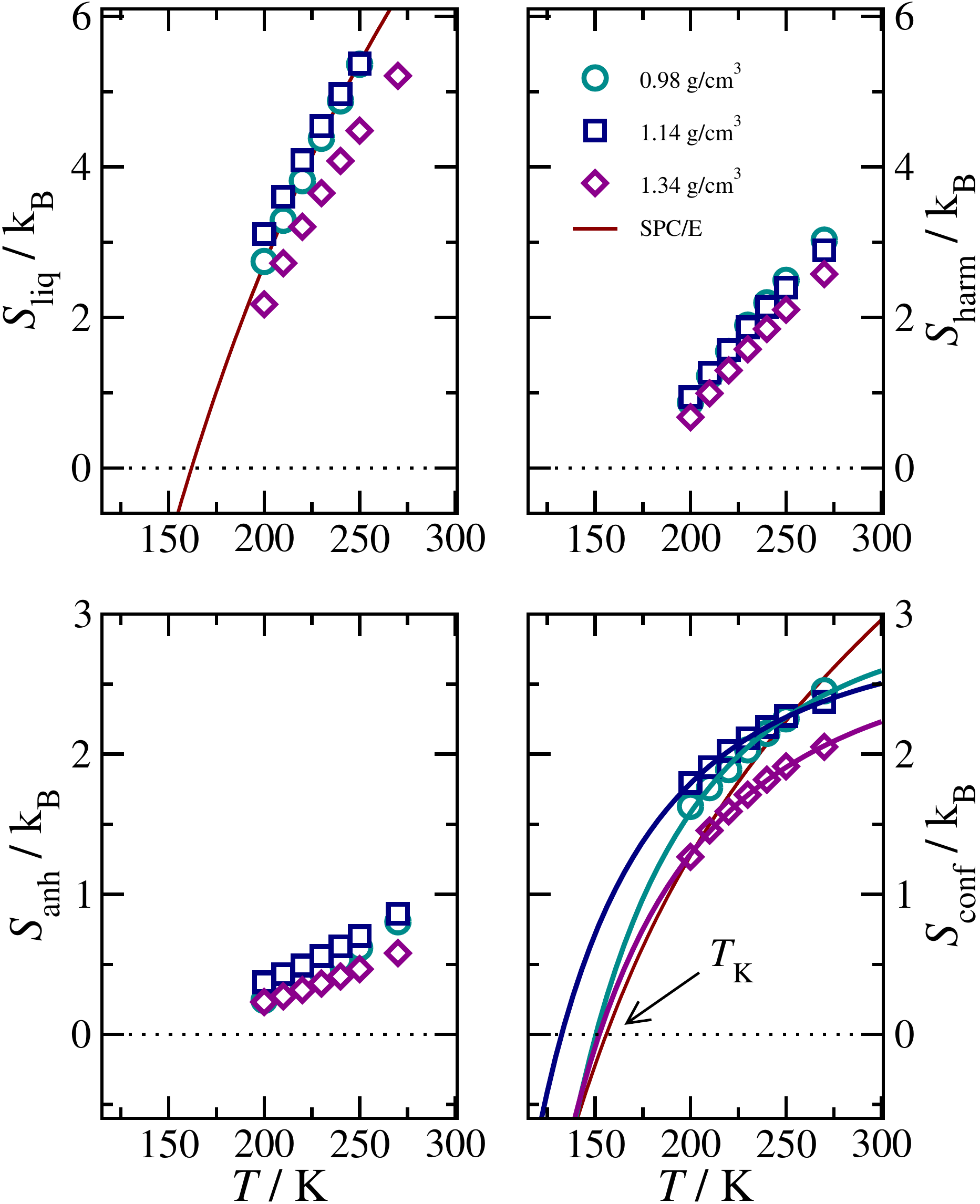}
\caption{Temperature dependence  of the entropy of the liquid $S_\text{liq}$, of the vibrational entropy (both harmonic $S_\text{harm}$ and anharmonic  $S_\text{anh}$ components)  and of the resulting configurational entropy $S_\text{conf}$ for $\rho=0.98$, $1.14$ and 1.34~g/cm$^{3}$ in comparison with result for SPC/E at $\rho=1.0$~g/cm$^{3}$ from Ref.~\cite{scala00-nature}. All quantities are expressed as per particle.}
\label{fig:entropy}
\end{figure}

 Figure~\ref{fig:entropy} shows the different entropic terms  and the resulting $S_\text{conf}$ for three different densities.
 Panel (d) also shows the PEL representation of $S_\text{conf}$ according to Eqn.~\ref{eq:s-fit}, with $\alpha$ as the only fit parameter. The PEL theoretical
 expression properly models the $T$ dependence of the numerical data.  The extrapolation of the theoretical curves toward $S_\text{conf}=0$ provides a visual
 estimate of the Kauzmann  $T$.   The resulting  density dependence   will be discussed in the next section together with the landscape phase diagram.
  The $\rho$ dependence of the best-fit values for $\alpha$ are shown in Figure~\ref{fig:params}(c).  $\alpha$ shows a monotonic dependence on $\rho$. The total number of basins,  $\exp(\alpha N)$ increases on decreasing density, with a trend consistent with what has been observed in all other studied models~\cite{sciortino05-jsm,mossa2002dynamics,sciortino03-prl}.

\subsection{The PEL-EOS}
\label{sec:peleos}

\begin{figure*}
  \includegraphics[width=1\textwidth]{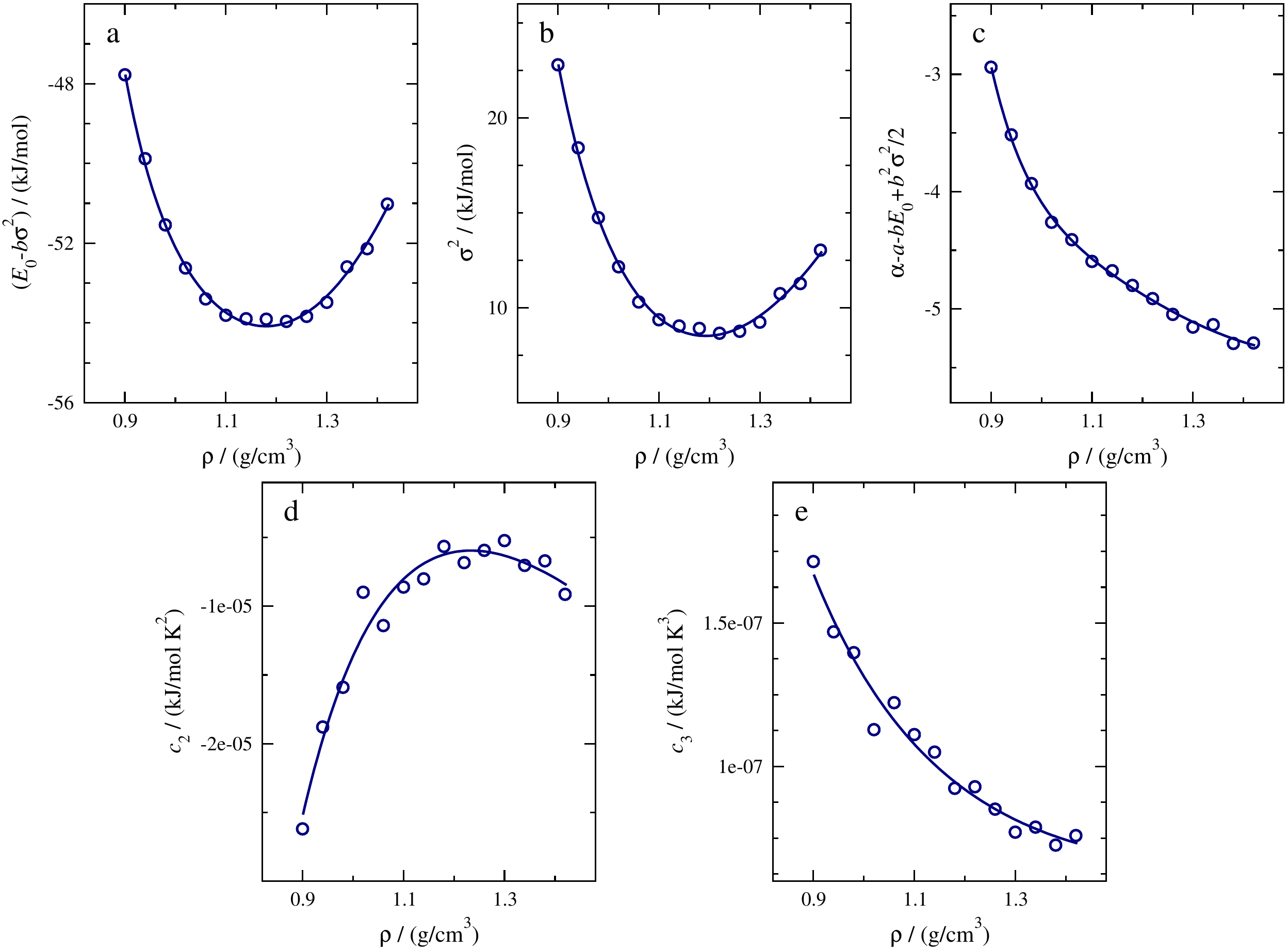}
\caption{Best polynomial fits  in $V$ of the quantities relevant for evaluating the PEL-EOS.   The polynomial fit has degree four
for the quantities in panels (a) (b) and (c) ($E_0-b\sigma^2$, $\sigma^2$ and $\alpha N -a-bE_0+b^2\sigma^2/2$, cf.
 Eqns. \ref{eq:ptm1}-\ref{eq:pt1}
) and degree two for the quantities in panels (d) and (e)  ($c_2$ and $c_3$, cf. Eqn. \ref{eq:pti}).
}
\label{fig:allpar-fit}
\end{figure*}

The $V$ derivatives of the landscape parameters allow us to  evaluate the PEL-EOS (see Eq.~\ref{eq:eos}). Fig.~\ref{fig:allpar-fit}
shows  the polynomial fits for
$\sigma^2$ (requested to evaluate $\mathcal{P}_{T^{-1}}$, Eq.~\ref{eq:ptm1}), for $E_0-b\sigma^2$ (to evaluate $\mathcal{P}_{T^{0}}$, Eq.~\ref{eq:pt0}) and
for $\alpha-a+bE_0+b^2\sigma^2/2$ (to evaluate $\mathcal{P}_{T^{1}}$, Eq.~\ref{eq:pt1}). The figure also shows  the 
anharmonic contribution $c_2$ and $c_3$. 
The functional form has been chosen as a four-degree polynomial in $V$ for $\sigma^2$, $E_0-b\sigma^2$ and $\alpha-a+bE_0+b^2\sigma^2/2$.
Due the the larger numerical error in the anharmonic contribution, we select a 
quadratic function in $V$ for $c_2$ and $c_3$. The resulting parameters are reported in  Appendix B for future reference.
   The selection of a fourth order in the  functional form generates a cubic $V$ dependence of the pressure,
which is the lowest order functional form consistent with the possibility of a liquid-liquid critical point. 

\begin{figure}
  \includegraphics[width=0.5\textwidth]{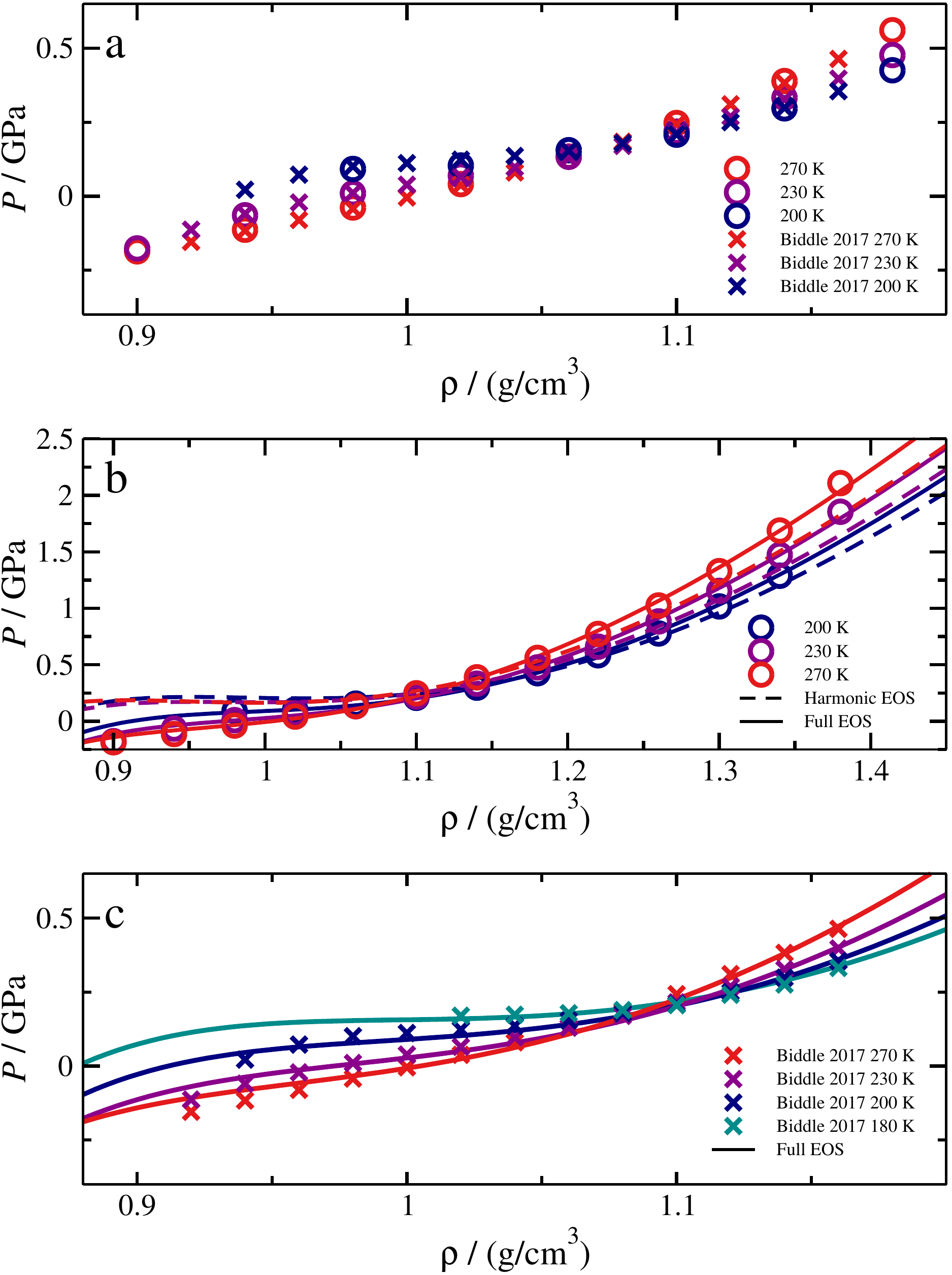}
\caption{Pressure-density
isotherms.  Panel (a) compares MD results from this study with recently published data   from the work of Biddle et al.~\cite{biddle17-jcp}.
Panel (b) compares the EOS resulting from the MD simulations with both the harmonic (dashed) and full (harmonic plus anharmonic, solid) landscape EOS. 
Panel (c) compares the EOS resulting from the MD simulations of Biddle et al.~\cite{biddle17-jcp} with the full PEL-EOS including the extrapolation of the 
theoretical PEL prediction to $T=180$~K, a $T$ which was not studied in the present work. 
}
\label{fig:conf-biddle}
\end{figure}

\begin{figure}
  \includegraphics[width=0.5\textwidth]{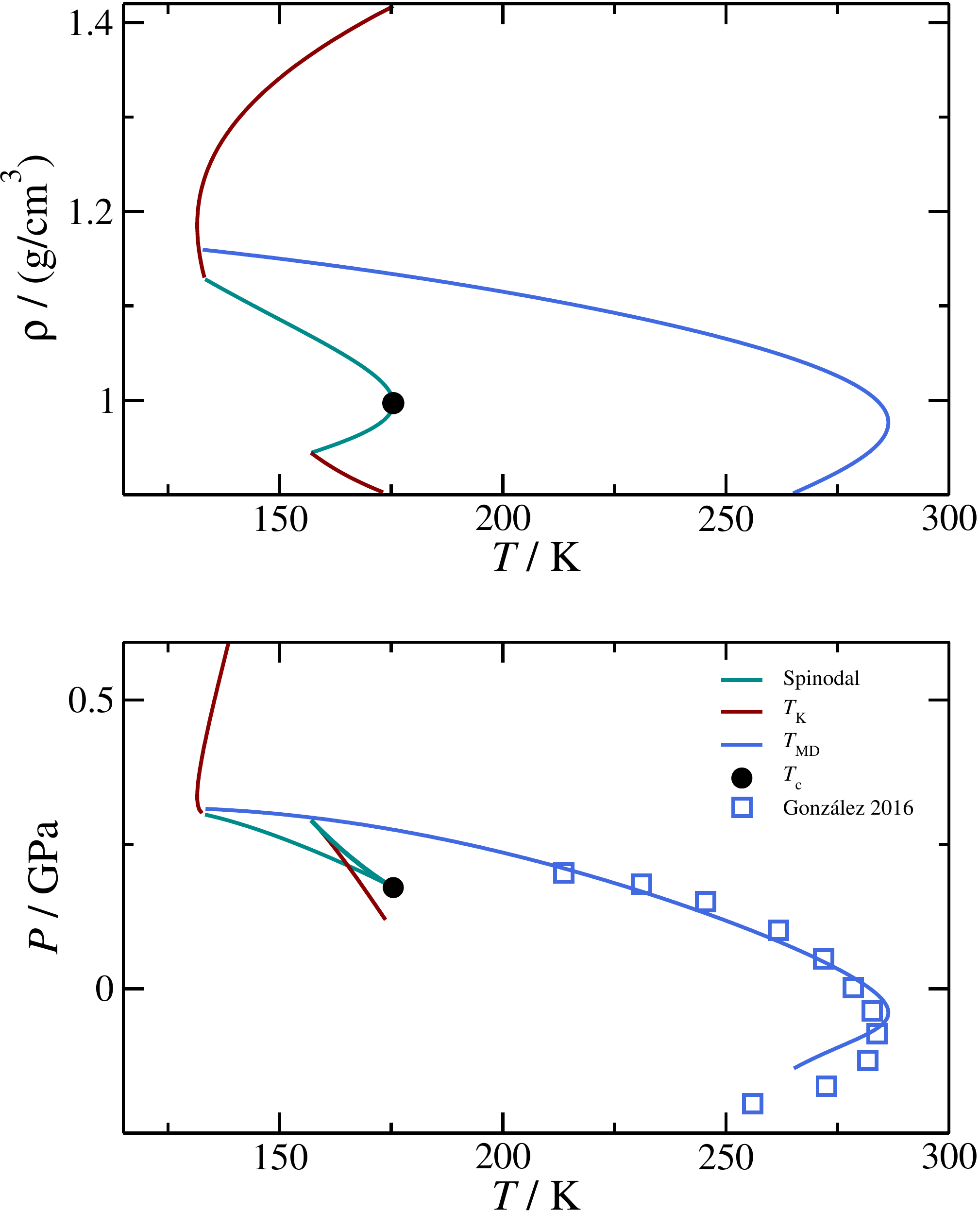}
\caption{Density-temperature (a) and pressure-temperature (b) PEL phase diagrams, reporting the liquid-liquid critical point and its associated mean-field spinodals, 
the Kauzmann  and the $T_\text{MD}$ loci.  It also shows in (b) the MD results for the  $T_\text{MD}$ locus from Ref.~\cite{gonzalez16-jcp}. 
}
\label{fig:pd}
\end{figure}

The resulting set of fitting coefficients thus allow us to rebuild the PEL-EOS for all $T$. The corresponding PEL-EOS for selected isotherms is shown in Fig.~\ref{fig:conf-biddle} and compared with the MD results at the same $T$. We also compare or MD data as well as the PEL-EOS with the previously published isotherms from  Biddle et al.~\cite{biddle17-jcp} in Fig.~\ref{fig:conf-biddle}. The PEL-EOS  approximates
the MD results rather well in the entire density range, stressing  the ability of a Gaussian landscape thermodynamic approach to  model 
the low $T$ behaviour of TIP4P/2005.  As expected from the quality of the comparison, but also from the $V$-dependence of $\sigma^2$,
the  PEL-EOS predicts the existence of a   temperature of maximum density ($T_\text{MD}$).  
The locus of the  $T_\text{MD}$ in the $P-T$ and in the $T-V$ plane  are shown in Fig.~\ref{fig:pd}, together with the MD results reported in Gonz\'{a}lez et al.~\cite{gonzalez16-jcp}. In the same figure we also show the projection of the Kauzmann locus, evaluated according to Eq.~\ref{eq:tk}. 
Within the PEL paradigm, this locus signals the limit of validity of the PEL-EOS previously derived. For lower $T$, the configurational entropy vanishes, the system
has reached the ideal-glass state (the  basin with energy $e_K$, see Eq.~\ref{eq:ek})  and the only residual contribution to the free-energy arises from the vibrational component.

The  derived PEL-EOS formally depend on  the assumption of a Gaussian PEL. In this respect, it offers a sound formulation of the
thermodynamics of the  TIP4P/2005 model which can be extended to temperatures below the lowest investigated one with the only
assumption that the landscape retains its Gaussian character.   It has been observed~\cite{sciortino03-prl} that a Gaussian PEL 
and a minimum in $\sigma^2$ vs. $V$ are the only ingredients requested to generate a low $T$ liquid-liquid critical point.
Consistent with this prediction, we determine for the  TIP4P/2005 model   the $T$ and $P$ values of 
the liquid-liquid critical point from the simultaneous vanishing of the first and second $V$ derivative of the EOS.  The best estimate is $T_\text{c}=175$~K,  $p_\text{c}=0.175$~GPa and $\rho_\text{c}=0.9970$~g/cm$^{3}$.  
The location of the predicted critical point is also shown in Fig.~\ref{fig:pd}.
We note that small variations of the  $\rho$-range included in the fit of the volume dependence of the landscape parameter do not significantly change
the predicted critical parameters. We then conclude that 
  the critical parameters are accurate within $T_\text{c}\pm2$~K,  $p_\text{c}\pm0.002$~GPa and $\rho_\text{c}\pm0.001$~g/cm$^{3}$.
The PEL estimate is consistent with the value of $T_\text{c}=193$ K reported in Ref.~\onlinecite{abascal10-jcp} as well as with the more recent estimate $T_\text{c}=182$ K~\cite{sumi13-rscadv,singh16-jcp,biddle17-jcp}.

\section{Conclusions}

In  this article we have reported a thorough analysis of the potential energy landscape statistical properties for the TIP4P/2005 model, one of the
most accurate classic models for water~\cite{vega11-pccp}.  To do so we evaluate the 
inherent structure and the local curvature around the IS for fourteen different  densities and seven different temperatures.

We have shown that   a Gaussian distribution of basin depth provides an accurate description of the system thermodynamics. 
 In the Gaussian landscape, just 
 three quantities ($E_0$, $\sigma^2$ and $\alpha$) control
 the thermodynamic behaviour of the system. 
 We have found that each molecule contributes to approximatively  $e^3$ basins, a number very close to the one 
which had been calculated for SPC/E previously.  
 From the volume derivative of these parameters, a landscape
 EOS has been derived.   Interestingly,  in harmonic approximation the $T$-dependence of the $P$ is condensed in three
 contributions,  proportional to $T^{-1}$, $T^{0}$ and $T^{1}$ respectively.  This simple $T$ dependence makes it possible to 
 identify immediately the PEL source of the density anomalies and the close connection between density anomalies
 and the existence of a  liquid-liquid critical point. Indeed, we have confirmed that for TIP4P/2005 
  the variance of the Gaussian distribution is found to display a
minimum as a function of the volume, the PEL signature of  densities anomalies.  

We have also shown that  including the anharmonic contributions makes it possible to accurately describe the PEL free-energy and
the the corresponding EOS.  The resulting  PEL-EOS  rather well approximates the MD $P(V,T)$, offering  a 
reliable functional form for predicting the low $T$ behaviour of the model.   A small $T$ extrapolation 
predicts a liquid-liquid critical point in  TIP4P/2005, consistent with previous estimates~\cite{sumi13-rscadv,singh16-jcp,biddle17-jcp}. 

Finally we note that
in the present investigation we have not revealed any signature of failure of   the Gaussian approximation.
This is rather well documented in Fig.~\ref{fig:eis-t} where a deviation from a linear
$1/T$ dependence is never observed. Still it is possible that, especially at low densities, where the system evolves toward a defect free
tetrahedral network and the bottom of the landscape (the fully bonded network) is approached,  the Gaussian approximation should
reveal its large-number origin and cease pace to a logarithmic landscape~\cite{debenedetti2003model}. Landscape analysis of the ST2 model~\cite{poole2005density}, of silica~\cite{saksaengwijit2004origin,saika2001fragile} and of  tetrahedral patchy particles~\cite{moreno2006non,smallenburg2013liquids}    suggest this possibility. Unfortunately, this cross-over temperature, if present for TIP4P/2005,
is still  below the $T$ range we have been able to explore with present day computational facilities.

\begin{acknowledgments}
PHH thanks the Austrian Science Fund FWF for support (Erwin Schr\"{o}dinger Fellowship J3811 N34). We also thank Chantal Valeriani for help in the initial stage of the project.  
\end{acknowledgments}

\appendix
\section{Free Energies of Supercooled TIP4P/2005}

In this section we report the molar free energies $F_\text{liq}$ of the TIP4P/2005 liquid at all studied state points  in the $T$ range $270-200$~K. The values  shown in Table~\ref{tab:free} were evaluated from the thermodynamic integration technique explained in section~\ref{sec:free}.

\begin{table*}

\caption{Molar free energies $F_\text{liq}$ in kJ~mol$^{-1}$ of the TIP4P/2005 liquid at given $T$ and $\rho$.}
\vskip 0.5cm
 \begin{tabular}{c|cccccccccccccc}
	
	 & \multicolumn{14}{c}{$\rho$~/~(g/cm$^{3}$)}\\
	$T$~/~K&	$0.9 $	&	$0.94 $	&	$0.98 $	&	$1.02 $&$1.06 $	&	$1.1 $	&	$1.14 $	&	$1.18 $	&	$1.22 $	&	$1.26 $	&	$1.3 $	&	$1.34 $	&	$ 1.38 $	&	$1.42$ 	\\
\hline

270	&	-56.91	&	-57.04	&	-57.10	&	-57.10	&	-57.04	&	-56.93&-56.74	&	-56.49	&	-56.16	&	-55.74	&	-55.23	&	-54.60	&	-53.87&-53.00	\\

250	&	-55.96	&	-56.08	&	-56.13	&	-56.12	&	-56.06	&	-55.95&-55.79	&	-55.55	&	-55.25	&	-54.86	&	-54.38	&	-53.80	&	-53.11&-52.29	\\

240	&	-55.56	&	-55.67	&	-55.70	&	-55.69	&	-55.63	&	-55.52&-55.36	&	-55.13	&	-54.84	&	-54.47	&	-54.00	&	-53.44	&	-52.77&-51.98	\\

230	&	-55.20	&	-55.30	&	-55.32	&	-55.29	&	-55.23	&	-55.12&-54.96	&	-54.75	&	-54.46	&	-54.10	&	-53.66	&	-53.12	&	-52.47&-	\\

220	&	-54.89	&	-54.98	&	-54.98	&	-54.94	&	-54.86	&	-54.76&-54.60	&	-54.40	&	-54.13	&	-53.78	&	-53.35	&	-52.84	&	-52.22&-	\\

210	&	-	&	-54.70	&	-54.68	&	-54.62	&	-54.54	&	-54.43&-54.28	&	-54.08	&	-53.83	&	-53.50	&	-53.09	&	-52.59	&	-	&-	\\

200	&	-	&	-	&	-54.41	&	-54.35	&	-54.26	&	-54.15&-54.01	&	-53.81	&	-53.56	&	-53.25	&	-52.86	&	-52.38	&	-	&-	\\
  \hline
 \end{tabular}
\label{tab:free}
\end{table*}

\section{PEL-EOS Parameters}

As explained in Section~\ref{sec:peleos} we fitted the volume dependence of the quantities relevant for the PEL-EOS (cf. Eqns.~\ref{eq:eos}-\ref{eq:pti}) by polynomials in molar volume $V$:
\begin{equation}
 f(V)=\sum_0^nA_iV^i,
\end{equation}
where $n$ was four for the  arguments in Eqns. \ref{eq:ptm1}-\ref{eq:pt1} and two for the $V$ dependence of $c_2$ and $c_3$ (cf. Eqn. \ref{eq:pti}). The resulting parameters, which suffice to build the PEL-EOS, are summarised in Table~\ref{tab:params}.

%

\begin{table*}
\caption{Fitting parameters of the polynomial fit in $V$ of the quantities ($Q$) relevant for the PEL-EOS. The unit of $A_i$ is the unit of $Q$ (cf. Fig.~\ref{fig:allpar-fit}) times the unit of inverse molar volume (mol/cm$^{3}$) to the $i$th power.}
\vskip 0.5cm
 \begin{tabular}{c|cccccc}
 Q & $A_0$ & $A_1$  & $A_2$ & $A_3$ & $A_4$\\
\hline

 $E_0-b\sigma^2$ &4.69736$\cdot 10^{2}$	&-1.19385$\cdot 10^{2}$	&1.02918$\cdot 10^{1}$	&-4.00425$\cdot 10^{-1}$	&5.98043$\cdot 10^{-3}$	\\
$\sigma^2$  & 5.50606$\cdot 10^{2}$	&-1.16995$\cdot 10^{2}$	&9.61032$\cdot 10^{0}$	&-3.63301$\cdot 10^{-1}$	&5.46496$\cdot 10^{-3}$	\\
$\alpha N -a-bE_0+b^2\sigma^2/2$ & 5.08848$\cdot 10^{1}$	&-1.53888$\cdot 10^{1}$	&1.55381$\cdot 10^{0}$	& -6.90207$\cdot 10^{-2}$	& 1.15370$\cdot 10^{-3}$	\\
$c_2$ & -1.46726$\cdot 10^{-4}$	& 1.92760$\cdot 10^{-5}$	&-6.59881$\cdot 10^{-7}$	&--	&--	\\
$c_3$ &1.53829$\cdot 10^{-7}$	&-1.84834$\cdot 10^{-8}$	&9.56831$\cdot 10^{-10}$	&--	&--	\\
 \end{tabular}
\label{tab:params}
\end{table*}

\bibliography{philip}

\end{document}